\documentclass[10pt,aps,prl,floatfix,showpacs,twocolumn]{revtex4-1}

\usepackage{graphicx}
\usepackage{amsmath, amsfonts, amssymb, bm,color}

\usepackage{amstext}
\usepackage{mathrsfs}

\newcommand{\corr}[1]{{\color{black}#1}}

\begin{document}

\title{Testing strong-field QED close to the fully non-perturbative regime using aligned crystals}
\author{A. Di Piazza}
\email{dipiazza@mpi-hd.mpg.de}
\author{T. N. Wistisen}
\author{M. Tamburini}

\affiliation{Max Planck Institute for Nuclear Physics, Saupfercheckweg 1, D-69117 Heidelberg, Germany} 

\author{U. I. Uggerh\o j}

\affiliation{Department of Physics and Astronomy, Aarhus University, Ny Munkegade 120, DK-8000 Aarhus, Denmark
}

\begin{abstract}
Processes occurring in the strong-field regime of QED are characterized by background electromagnetic fields of the order of the critical field $F_{cr}=m^2c^3/\hbar|e|$ in the rest frame of participating charges. It has been conjectured that if in their rest frame electrons/positrons experience field strengths of the order of $F_{cr}/\alpha^{3/2}\approx 1600\,F_{cr}$, with $\alpha\approx 1/137$ being the fine-structure constant, their effective coupling with radiation becomes of the order of unity. Here we show that channeling radiation by ultrarelativistic electrons with energies of the order of a few TeV on thin tungsten crystals allows to test the predictions of QED close to this fully non-perturbative regime by measuring the angularly resolved single photon intensity spectrum. The proposed setup features the unique characteristics that essentially all electrons 1) undergo at most a single photon emission and 2) experience at the moment of emission and in the angular region of interest the maximum allowed value of the field strength, which at $2\;\text{TeV}$ exceeds $F_{cr}$ by more than two orders of magnitudes in their rest frame.
\end{abstract}

\maketitle

A measure of the strength of the electromagnetic interaction is theoretically represented by the dimensionless fine-structure constant $\alpha=e^2/\hbar c$, with $e<0$ being the electron charge and in units where $4\pi\epsilon_0=1$ \cite{Jauch_b_1976,Itzykson_b_1980,Landau_b_4_1982,Schwartz_b_2014}. At energies of the order of the electron rest energy $mc^2\approx 0.511\;\text{MeV}$, the numerical value of the fine-structure constant is about $1/137\approx 7\times 10^{-3}$. It is known, however, that at increasingly high energies the effective electromagnetic coupling becomes increasingly larger and features a pole (Landau pole) at $\Lambda_{\text{QED}}\sim mc^2\exp(3\pi/2\alpha)\sim 10^{277}\;\text{GeV}$ \cite{Jauch_b_1976,Itzykson_b_1980,Landau_b_4_1982,Schwartz_b_2014}. From a pragmatic point of view the Landau pole does not limit the applicability of QED and its exceedingly large value is closely related to the fact that radiative corrections in QED become larger only logarithmically at increasingly high energies \cite{Jauch_b_1976,Itzykson_b_1980,Landau_b_4_1982,Schwartz_b_2014}.

In view of the experimental success of QED, it is natural to test the theory under extreme conditions, such as those provided by intense background electromagnetic fields. In the realm of QED, a background field is denoted as ``intense'' if the probabilities of quantum processes show a significant nonlinear dependence on the field amplitude. In this regime the background electromagnetic field must be taken into account exactly in the calculations, which is achieved by quantizing the electron-positron field in the presence of the field (Furry picture) \cite{Furry_1951}. Additionally, quantum effects like photon recoil significantly alter the probabilities of quantum processes if electrons and positrons experience in their rest frame electromagnetic fields of the order of the ``critical'' fields of QED: $E_{cr}=m^2c^3/\hbar|e|\approx 1.3\times 10^{16}\;\text{V/cm}$ and $B_{cr}=m^2c^3/\hbar|e|\approx 4.4\times 10^{13}\;\text{G}$ \cite{Landau_b_4_1982,Dittrich_b_1985,Ritus_1985,Fradkin_b_1991,Ehlotzky_2009,Di_Piazza_2012,Dunne_2014}. \corr{At background electric fields of the order of $E_{cr}$ the vacuum becomes unstable under electron-positron pair production and at background magnetic fields of the order of $B_{cr}$ the magnetic energy related to the electron magnetic moment becomes comparable with $mc^2$. The instability of the vacuum in an ultra-critical magnetic field due to the collapse of positronium has been investigated in Refs. \cite{Shabad_2006a,Shabad_2006b}}.

Ultrarelativistic electrons entering a single crystal (almost) along a direction of high symmetry (below denoted as $z$) interact coherently with the crystal atoms aligned along the symmetry direction  \cite{Akhiezer_1993,Akhiezer_b_1996,Baier_b_1998,Rullhusen_b_1998,Uggerhoj_2005,Andersen_2012,Di_Piazza_2017,Wistisen_2018,Wistisen_2019,Khokonov_2019}. In this regime, the electric field of all aligned atoms can be approximately described by means of a continuous potential and the total crystal field is the sum of the electric fields of all the ``strings'' of atoms periodically distributed on the transverse ($xy$) plane according to the structure of the crystal. If an electron with initial energy $\varepsilon\gg mc^2$ enters a crystal with a velocity at an angle $\theta\ll 1$ with respect to the $z$ axis, the electron motion in the continuous potential becomes transversely bound if $\theta\lesssim\theta_c=\sqrt{2|U_M|/\varepsilon}$ (axial channeling) \cite{Lindhard_1965}, where $U_M$ is the electron potential energy depth in the crystal. In the case of channeling, nonlinear effects become sizable if the motion on the $xy$ plane is relativistic, i.e., if $\xi=\varepsilon\theta_c/m=\sqrt{2|U_M|\varepsilon/m^2}\gtrsim 1$ (from now on units with $\hbar=c=1$ are employed) \cite{Akhiezer_b_1996,Baier_b_1998,Rullhusen_b_1998,Uggerhoj_2005}. Quantum effects like photon recoil are instead controlled by the quantum nonlinearity parameter $\chi=(\varepsilon/m)E_{\perp}/E_{cr}$, where $E_{\perp}$ is a measure of the crystal field on the $xy$ plane \cite{Akhiezer_b_1996,Baier_b_1998,Rullhusen_b_1998,Uggerhoj_2005}.

In the seventies Ritus and Narozhny conjectured that at $\chi\gg 1$ the effective coupling of QED in a constant crossed field (CCF), i.e., a constant and uniform electromagnetic field $(\bm{E}_0,\bm{B}_0)$ such that the two Lorentz-invariant quantities $\bm{E}_0^2-\bm{B}_0^2$ and $\bm{E}_0\cdot\bm{B}_0$ vanish, scales as $\alpha\chi^{2/3}$ \cite{Ritus_1970,Narozhny_1979,Narozhny_1980,Morozov_1981} (see also Ref. \cite{Akhmedov_1983} and the reviews in Refs. \cite{Ritus_1985,Akhmedov_2011,Fedotov_2017}). Since, apart from inessential prefactors, the energy of the incoming particle enters radiative corrections only through $\chi$ at $\chi\gg 1$, the Ritus-Narozhny (RN) conjecture implies an asymptotic high-energy behavior of strong-field QED in a CCF qualitatively different from that of QED in vacuum (see also Refs.~\cite{Podszus_2019,Ilderton_2019} for an analysis about the interplay between the high-energy limit and the CCF limit in strong-field QED). It has been recently shown that this fully non-perturbative regime of strong-field QED can be entered by employing intense laser radiation \cite{Blackburn_2019,Baumann_2019} and collision between dense electron and/or positron bunch collisions \cite{Yakimenko_2019,Tamburini_2019}. Both from an experimental and a theoretical point of view, however, it is crucial to identify a physical observable that can be measured and computed, such that strong-field QED can be effectively put to the test.

In the present Letter we show that channeling radiation by electrons with a few TeV energy on thin tungsten crystals represents a promising tool to approach this extreme regime of QED and to test the theory by measuring the angularly-resolved single-photon intensity spectrum. Indeed, the present setup has the unique features that essentially all electrons undergo at most a single photon emission and, in the angular region of interest, emit at the the maximum allowed value of the parameter $\chi\gtrsim 100$, which allows for a feasible comparison between experimental results and theoretical predictions. Now, at the CERN Secondary Beam Areas (SBA) beamlines electron beams of energies up to about $250\;\text{GeV}$ are available and it has already been proposed to extract the proton beam from the Large Hadron Collider (LHC) to produce secondary electrons with energies up to about $4\;\text{TeV}$ \cite{Uggerhoj_2005b,Massacrier_2018}, which in tungsten can experience fields corresponding to $\chi\gtrsim 300$ \cite{Uggerhoj_2005b}. We point out that our aim here is not to \emph{enter} the regime where $\alpha\chi^{2/3}\sim 1$, which corresponds to $\chi\sim 1600$, because a fully non-perturbative theory of strong-field QED is not available yet. We rather aim at values of $\chi\gtrsim 100$ such that $\alpha\chi^{2/3}$ is significantly larger than $\alpha$ but still sufficiently smaller than unity that a perturbative treatment of the interaction between electrons/positrons and radiation field based on the Furry picture is applicable.

Below we consider a bunch of ultrarelativistic electrons impinging one by one onto a tungsten crystal (approximately) along the $\langle 111\rangle$ direction, which is set to coincide with the $z$ direction of the coordinate system. First, we present some analytical considerations on the electric field of a single atomic string. \corr{Note that by estimating the variation $\Delta\rho$ of the impact parameter between two successive atoms in the string as $\Delta\rho\sim d\theta$, where $d$ is the atomic spacing and $\theta\sim 2Z\alpha/\varepsilon\rho$ is the typical deflection angle for an ultrarelativistic electron ($Z=74$ for tungsten), we conclude that the continuous-potential model is applicable in the ultrarelativistic regime because the condition $\Delta\rho\ll\rho$ is fulfilled at $\rho\sim R$, where $R\approx \lambda_C/\alpha Z^{1/3}$ is the Thomas-Fermi radius, with $\lambda_C=1/m\approx 3.9\times 10^{-11}\;\text{cm}$ being the Compton wavelength \cite{Akhiezer_b_1996}.}

By indicating as $\bm{\rho}=(x,y)$ the coordinates in the transverse plane, with the atomic string crossing this plane at $\bm{\rho}=\bm{0}$, the continuum potential $\Phi(\rho)$ depends only on the distance $\rho=|\bm{\rho}|$ and it can be approximated 
as \cite{Baier_b_1998}:
\begin{equation}
\label{Phi}
\Phi(\rho)=
\Phi_{0}\left[\ln\left(1+\frac{1}{\varrho^2+\eta}\right)-\ln\left(1+\frac{1}{\varrho_c^2+\eta}\right)\right],
\end{equation}
where $\bm{\varrho}=\bm{\rho}/a_s$ and $\varrho_c=\rho_c/a_s$. Here, the parameters $\Phi_0$, $\rho_c$, $\eta$, and $a_s$ depend on the crystal and $\rho\le\rho_c$. In the case of the $\langle 111\rangle$ axis of tungsten we have that $\Phi_0=417\;\text{V}$, $\rho_c=1.35\;\text{\AA}$, $\eta=0.115$, and $a_s=0.215\;\text{\AA}$. The physical meaning of these parameters is clear from Eq. (\ref{Phi}) (the parameter $U_M$ introduced above corresponds here to $e\Phi(0)$) and we only point out that the quantity $\pi\rho_c^2$ represents the area per unit string corresponding here to the $\langle 111\rangle$ direction in tungsten \cite{Baier_b_1998}. The electric field vector $\bm{E}_{\perp}(\bm{\rho})$ of the string lies on the $xy$ plane and is given by (see also Fig. 1)
\begin{equation}
\bm{E}_{\perp}(\bm{\rho})=-\bm{\nabla}_{\perp}\Phi(\rho)=\frac{2\Phi_0}{a_s}\frac{\bm{\varrho}}{\eta+\varrho^2+(\eta+\varrho^2)^2}.
\end{equation}
\begin{figure}
\begin{center}
\includegraphics[width=\columnwidth]{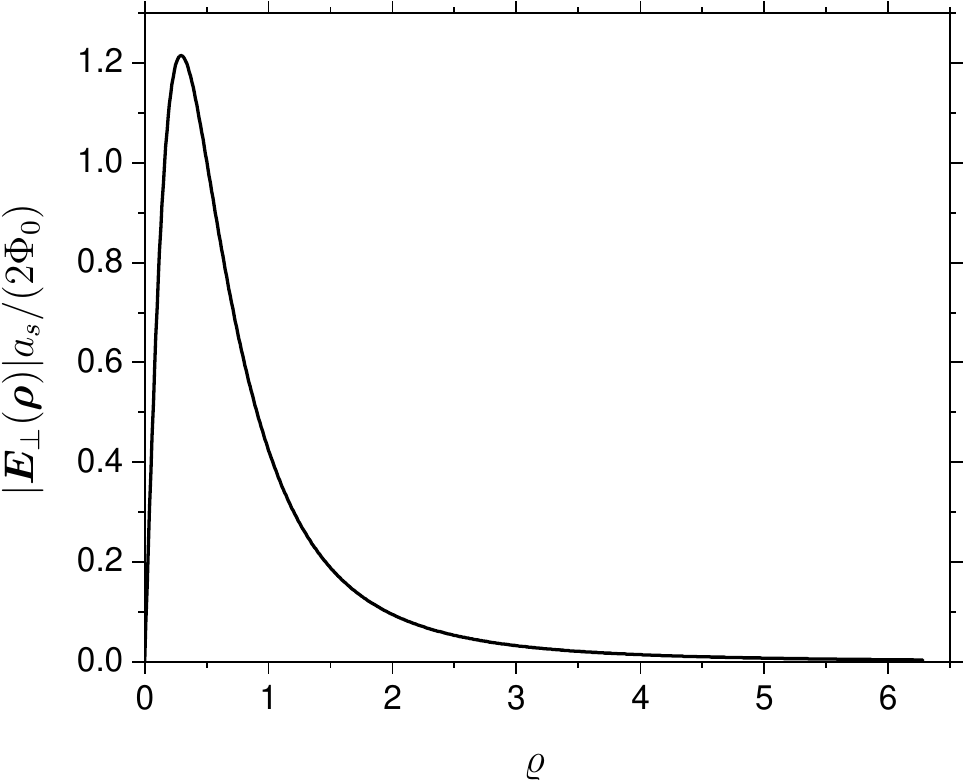}
\caption{Amplitude of the electric field of an atomic string of tungsten along the $\langle 111\rangle$ direction in units of $2\Phi_0/a_s\approx 3.88\times 10^{11}\;\text{V/cm}$ as a function of $\varrho=\rho/a_s$ for $0\le\varrho\le \rho_c/a_s\approx 6.28$.}
\end{center}
\end{figure}

The corresponding (local) quantum nonlinearity parameter reads
\begin{equation}
\chi(\rho)=\frac{\varepsilon}{m}\frac{|e\bm{E}_{\perp}(\bm{\rho})|}{m^2}=\frac{2\varepsilon|U_0|}{m^2}\frac{\lambda_C}{a_s}\frac{\varrho}{\eta+\varrho^2+(\eta+\varrho^2)^2},
\end{equation}
where $U_0=e\Phi_0$ and where we have implicitly assumed that the electron energy $\varepsilon$ does not significantly change during the interaction with the crystal (see also below). In order to interpret the numerical results reported below, it is useful to notice that $\chi(\rho)$ has a maximum approximately at $\rho=\rho_{\text{max}}\approx\sqrt{\eta}(1-2\eta)a_s\approx 0.261\,a_s$ (see also Fig. 1) and that
\begin{equation}
\label{chi_max}
\chi_{\text{max}}=\chi(\rho_{\text{max}})\approx\frac{\varepsilon|U_0|}{m^2}\frac{\lambda_C}{a_s}\frac{1-2\eta}{\sqrt{\eta}}\approx 65.8\,\varepsilon[\text{TeV}].
\end{equation}

Now, we report the results of numerical simulations in which an electron enters a tungsten crystal of \corr{$L=5\;\text{$\mu$m}$} thickness (almost) along the $\langle 111\rangle$ direction. Although the above analytical considerations are restricted to a single atomic string, the presented numerical results are obtained by determining the crystal field as the sum of the contributions of the 41 strings closest to the electron on the transverse $xy$ plane. For the potential of each string, the Doyle-Turner potential has been employed \cite{Doyle_1968,Peng_1999}, which is more accurate than that in Eq. (\ref{Phi}) but less suitable for analytical considerations. The results are obtained by averaging over a bunch of $2\times 10^7$ electrons. The electrons are uniformly distributed on the $xy$ plane, have an energy of $2\;\text{TeV}$ and initially Gaussian distributed opening angles along the $x$ and the $y$ direction both centered around zero and with standard deviation of $5\;\text{$\mu$rad}$, which is not an unrealistic scaling with energy of what can be achieved nowadays \cite{Wistisen_2018}. \corr{Now, in the ultrarelativistic regime under consideration the photon emission probability can be computed within the semiclassical method \cite{Baier_b_1998}, which requires the knowledge of the electron classical trajectory in the crystal field. Thus,} the numerical code computes the evolution of the electrons inside the crystal via the Lorentz equation \corr{ (for the sake of the estimate, note that the number of bound states in the transverse motion within the single-string model is of the order of $(a_s/\lambda_C)\sqrt{\varepsilon|U_0|/m^2}\sim  10^3\gg 1$ \cite{Baier_b_1998,Di_Piazza_2017}, such that their discrete nature can be ignored)}. The emission of photons is implemented by means of a Monte Carlo algorithm, with the emission probabilities per unit time given by the corresponding expressions within the local constant field approximation (see, e.g., Eq. (4.24) in Ref.~\cite{Baier_b_1998}). Analogously, any emitted photon may decay into an electron positron pair (see, e.g., Eq. (3.50) in Ref.~\cite{Baier_b_1998} for the corresponding probability per unit time) although, due to the short thickness of the crystal this process turned out to be negligible. Moreover, and for the same reasons, the total probability of photon emission was much smaller than unity in the simulations. \corr{Also, an energy $\mathcal{E}_t=\alpha(\omega_p/m)\varepsilon/3\approx 3.9\times 10^{-7}\,\varepsilon$, with $\omega_p\approx 1.6\times 10^{-4}\,m$ being the tungsten plasma frequency, is emitted as transition radiation mostly at photon energies $\omega_t\lesssim (\omega_p/m)\varepsilon\approx 1.6\times 10^{-4}\,\varepsilon$ \cite{Jackson_b_1975}, such that it can be safely neglected.} It is also worth mentioning that it is appropriate to use here the probabilities within the local field approximation because $\xi\approx 85$ and, as we will see below, $\chi\lesssim 200$ such that $\xi^3\gg \chi$ (recall that each electron essentially emits at most one photon) \cite{Baier_b_1998,Di_Piazza_2007,Dinu_2016,Di_Piazza_2018c,Podszus_2019}. In addition, the finite extension of the crystal does not prevent the use of the CCF approximation at $\omega\sim \varepsilon$ \cite{Wistisen_2015} because, e.g., at large values of $\chi$ the formation length $l_f$ of a photon with energy $\omega$ is $l_f\sim\varepsilon\lambda_C[24(\varepsilon-\omega)/\omega]^{1/3}/m\chi^{2/3}$ and at $\omega\sim \varepsilon$, i.e., at $(\varepsilon-\omega)/\omega\sim 1$, and $\chi\sim 200$ it is $l_f\sim 0.1\;\text{$\mu$m}\ll L$ \cite{Baier_b_1998}.
\begin{figure}
\begin{center}
\includegraphics[width=\columnwidth]{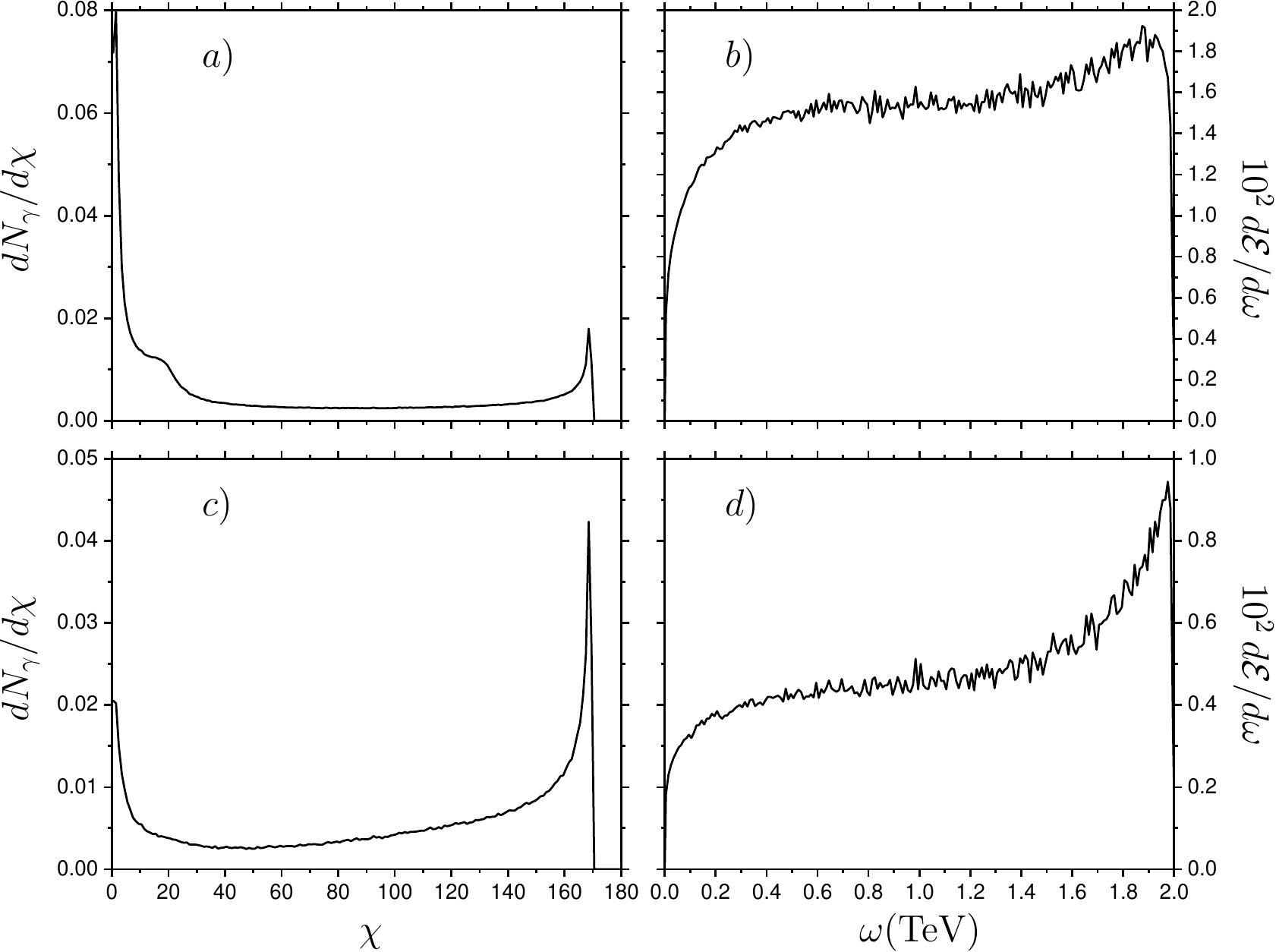}
\caption{Left panels: Emitted photon number distributions (normalized to unity) as functions of the electron quantum nonlinearity parameter at the moment of emission, accounting for all emitted photons (Fig. 2a) and for the photons emitted with an angle larger than $\theta_c/2$ with respect to the $z$ axis (Fig. 2c). Right panels: Average energy emitted per unit of photon energy, accounting for all emitted photons (Fig. 2b) and for the photons emitted with an angle larger than $\theta_c/2$ with respect to the $z$ axis (Fig. 2d). The other parameters of the crystal and of the electron bunch are reported in the text.}
\end{center}
\end{figure}

In Fig. 2a we report the distribution $dN_{\gamma}/d\chi$ normalized to unity, where $dN_{\gamma}=(dN_{\gamma}/d\chi)d\chi$ is the total number of photons emitted by all electrons with quantum parameter between $\chi$ and $\chi+d\chi$. Also, Fig. 2b shows the corresponding photon intensity spectrum, i.e., the average energy emitted per unit of photon energy. Figure 2a shows that most of the photons are emitted at low values of $\chi$, as it is physically expected because the oscillating electrons spend more time far from the strings where the field is relatively small. However, a peak is visible corresponding to the maximum allowed value of $\chi$, which is in good agreement with the value predicted by Eq. (\ref{chi_max}). The presence of the peak at $\chi_{\text{max}}$ can be explained by employing the single-string model. In fact, the number of photons $dn_{\gamma}(\chi,t)$ emitted between the times $t$ and $t+dt$ with the quantum nonlinearity parameter between $\chi$ and $\chi+d\chi$ is given by the emission probability $(dP_{\gamma}(\chi,t)/dt)dt$ times the number $dn_e(\chi,t)$ of electrons emitting between $t$ and $t+dt$ at a value of the quantum nonlinearity parameter between $\chi$ and $\chi+d\chi$. Now, within the single-string model the parameter $\chi$ is a function of the transverse distance $\rho$ from the string. However, since the field of the string vanishes at $\rho=0$ and (almost) at $\rho=\rho_c$ featuring a maximum at $\rho_{\text{max}}$ (see also Fig. 1), two values $\rho_+$ and $\rho_-$ of $\rho$ correspond to any value of $\chi\in(\chi(\rho_c),\chi_{\text{max}})$ such that $d\chi(\rho_{\pm})/d\rho\gtrless 0$. By limiting to this range of values of $\chi$ for the sake of simplicity (note that $\chi(\rho_c)\approx 1.8\times 10^{-3}\chi_{\text{max}}$), we can conclude that the total number $dN_{\gamma}(\chi)/d\chi$ of photons emitted per units of $\chi$ is given by
\begin{equation}
\label{dndchi}
\begin{split}
\frac{dN_{\gamma}}{d\chi}&=\int_0^Ldt \frac{dP_{\gamma}(\chi,t)}{dt}\sum_{i=\pm}\frac{d\tilde{n}_e(\rho_i,t)}{d\rho}\left(\frac{d\chi(\rho_i)}{d\rho}\right)^{-1}.
\end{split}
\end{equation}
This equation clearly explains the appearance of the peak at $\chi_{\text{max}}$ in the quantity $dN_{\gamma}/d\chi$ because the function $d\tilde{n}_e(\rho,t)/d\rho=dn_e(\chi(\rho),t)/d\rho$ is a smooth function of $\rho$ (at $t=0$ it corresponds to the uniform distribution of the electrons on the $xy$ plane) and $d\chi(\rho)/d\rho$ vanishes when $\chi(\rho)$ reaches its maximum.

Now, as we have observed, the spectrum in Fig. 2a receives contributions from emissions occurring both at high and at low values of $\chi$ and our aim is to identify an observable quantity, which only stems from emissions occurring at large values of $\chi$. In order to isolate the contribution of the photons emitted at high values of $\chi$, we make the following considerations again based on the single-string model and, for the sake of simplicity, we ignore the initial electron transverse velocity. Due to their uniform distribution, most of the electrons have initially a relatively large value of $\rho$. Now, these electrons can emit at the maximum value of $\chi$ only when they cross the region of largest field, which is relatively close to the string (see Fig. 1). Thus, the emissions at large $\chi$ by these electrons occur when the angle between the velocity of these electrons and the $z$ axis is close to the critical angle $\theta_c$. Correspondingly, the photons are essentially emitted with angles with respect to the $z$ axis of that order of magnitude. The idea is then to isolate the emissions at high $\chi$ by detecting only photons emitted at relatively large angles. We have exploited this idea and, in Figs. 2c and 2d, decided to consider only photons emitted at angles larger than $\theta_c/2$, with the numerical value of the potential depth $U_M$ of the Doyle-Turner potential (recall that $\theta_c=\sqrt{2|U_M|/\varepsilon}$). Figure 2c clearly shows that indeed most of the photons emitted at large angles are also emitted at the maximum value of $\chi$. For the sake of a clear visualization, both distributions in Figs. 2a and 2c have been normalized to unity. Despite the normalization, we can conclude that whereas in Fig. 2a the peak at high $\chi$ was about four times lower than that at low $\chi$, by cutting the ``straight'' photons we obtain that the peak at high $\chi$ is about two times higher than the peak at low $\chi$. Correspondingly the spectrum in Fig. 2d is much ``harder'' than that Fig. 2b. Also, we point out that by computing the total number of emitted photons (in any direction), we found that on average each electron emits $0.1$ photons and that the process of pair production by these photons is safely negligible. This is an important virtue of the present setup, which guarantees that the occurrence of multiple photon emissions and the development of electromagnetic showers can be ignored. Thus, the spectra in Figs. 2b and 2d are essentially single-photon spectra that can be correspondingly computed theoretically by employing the Furry picture.

Now, in order to confirm more quantitatively that indeed the spectrum Fig. 2d arises from photons emitted at large values of $\chi$, we first normalize the spectrum to unity, then we rescale the photon energies to the initial electron energy, and we compare the resulting spectrum with the ``universal'' normalized expression of the single-photon intensity spectrum $d\mathcal{E}_n/d\omega$ for large values of $\chi$. It is worth reporting this expression, which, appropriately normalized to unity, reads \cite{Baier_b_1998}
\begin{equation}
\label{dE_domega}
\frac{d\mathcal{E}_n}{d\omega}=\frac{81\sqrt{3}}{64\pi}\left(\frac{\omega}{\varepsilon-\omega}\right)^{1/3}\frac{\varepsilon^2+(\varepsilon-\omega)^2}{\varepsilon^2}.
\end{equation}
This expression is valid for $\chi\gg \omega/(\varepsilon-\omega)$, i.e., it becomes inapplicable only at the high-energy end of the spectrum where the spectrum goes exponentially to zero (at $\omega/(\varepsilon-\omega)\gg \chi$). However, if one integrates the asymptotic expression of the differential intensity of radiation, which corresponds to Eq. (\ref{dE_domega}) when normalized to unity, one already obtains the correct asymptotic of the total intensity of radiation with the scaling $\alpha\chi^{2/3}$ (see Eq. (4.28) in \cite{Baier_b_1998}). In Fig. 3 we compare the normalized intensity spectrum obtained with the angular cut (black solid curve) with $d\mathcal{E}_n/d\omega$ (red dashed curve), and we can see that the agreement is good except at the end of the spectrum where $d\mathcal{E}_n/d\omega$ formally diverges (and then becomes inapplicable). 
\begin{figure}
\begin{center}
\includegraphics[width=\columnwidth]{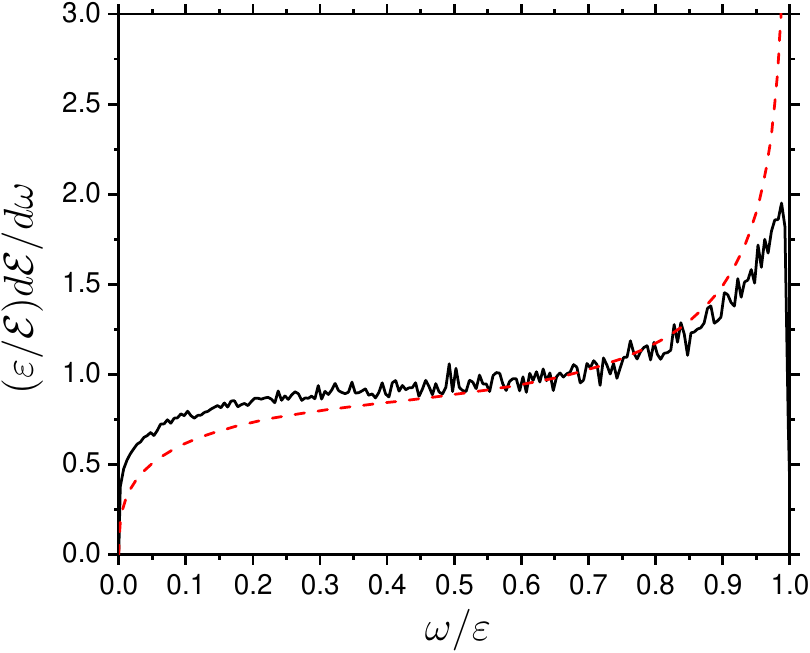}
\caption{Normalized intensity of radiation as a function of the emitted photon energy in units of the initial electron energy as obtained from the spectrum in Fig. 2d (black solid curve) and from the analytical expression in Eq. (\ref{dE_domega}) (red dashed curve). See the text for the electron and the crystal numerical parameters.}
\end{center}
\end{figure}

Finally, we point out that the value of $\chi_{\text{max}}\sim 100$ in the present setup is close to optimal in the following sense. At the present time to make theoretical predictions at such large values of $\chi$ that $\alpha\chi^{2/3}\approx 1$ is not possible because in principle all higher-order radiative corrections have to be taken into account in any calculation. Instead, Fig. 2c indicates that although $\chi_{\text{max}}$ is much larger than unity ($\approx 170$), we have $\alpha\chi_{\text{max}}^{2/3}\approx 0.2$. In this way, although we have computed the emission spectra only at the leading order and by ignoring radiative corrections, we expect that corresponding exact QED emission spectra (and then also the experimental spectra) would differ from the results here by $\sim 20\%$ (according to the preliminary estimations in Ref. \cite{Yakimenko_2019}, smaller corrections might be expected). A complete theoretical analysis of the leading-order radiative corrections, however, is beyond the scope of the present paper.

In conclusion, we have demonstrated that by employing strong-field radiation by electrons of a few-TeV energy crossing a thin tungsten target, the fully non-perturbative QED regime can in principle be approached and a clear physical observable can be identified in the angularly-resolved single-photon intensity spectrum. High-energy electron beams with energies of the order of a few TeV can in principle be obtained at CERN by extracting the proton beam from the LHC, which would render this regime soon accessible experimentally for the first time.

\begin{acknowledgments}
\corr{TNW was supported by the Alexander von Humboldt-Stiftung. ADP and MT
acknowledge useful discussions with Sebastian Meuren.

MT was the first to investigate the supercritical regime of QED
back in 2014 and found intriguing results in a different setup. 
ADP has then conceived the ideas of testing the supercritical
regime of QED with crystals and of employing the angularly-resolved
spectrum as observable, with inputs and discussions from all coauthors,
especially from TNW. TNW has proposed to compare the spectrum
with the corresponding asymptotic theoretical expression. UIU has provided the input 
about the possible experimental realization of the setup, the extraction of the 
proton beam from the LHC and the features of the resulting secondary 
beam. The numerical simulations were carried out by ADP and TNW. 
The manuscript was written by ADP with input from all coauthors.}
\end{acknowledgments}

%merlin.mbs apsrev4-1.bst 2010-07-25 4.21a (PWD, AO, DPC) hacked
%Control: key (0)
%Control: author (8) initials jnrlst
%Control: editor formatted (1) identically to author
%Control: production of article title (-1) disabled
%Control: page (0) single
%Control: year (1) truncated
%Control: production of eprint (0) enabled
%

%\bibliography{/home/theo/tonywolf/Samba/Travagghiu/Bibliography/arXiv,/home/theo/tonywolf/Samba/Travagghiu/Bibliography/Books,/home/theo/tonywolf/Samba/Travagghiu/Bibliography/Reviews,/home/theo/tonywolf/Samba/Travagghiu/Bibliography/Papers_Nuclei,/home/theo/tonywolf/Samba/Travagghiu/Bibliography/Papers_Radiation,/home/theo/tonywolf/Samba/Travagghiu/Bibliography/Papers_RR,/home/theo/tonywolf/Samba/Travagghiu/Bibliography/Papers_PP_and_Cascades,/home/theo/tonywolf/Samba/Travagghiu/Bibliography/Papers_VPE,/home/theo/tonywolf/Samba/Travagghiu/Bibliography/Papers_Crystal,/home/theo/tonywolf/Samba/Travagghiu/Bibliography/Papers_Various,/home/theo/tonywolf/Samba/Travagghiu/Bibliography/Papers_Laser_Plasma_QED}
%\bibliography{g://Travagghiu/Bibliography/arXiv,g://Travagghiu/Bibliography/Books,g://Travagghiu/Bibliography/Reviews,g://Travagghiu/Bibliography/Papers_Nuclei,g://Travagghiu/Bibliography/Papers_Radiation,g://Travagghiu/Bibliography/Papers_RR,g://Travagghiu/Bibliography/Papers_PP_and_Cascades,g://Travagghiu/Bibliography/Papers_VPE,g://Travagghiu/Bibliography/Papers_Crystal,g://Travagghiu/Bibliography/Papers_Various,g://Travagghiu/Bibliography/Papers_Laser_Plasma_QED}

\end{document}